\documentclass[aps,prb,twocolumn,notitlepage,groupedaddress,showpacs]{revtex4-1}

\usepackage{graphicx}

\begin{document}


\title{Magnetic anisotropy energy of disordered tetragonal
       Fe-Co systems from ab initio alloy theory}


\author{I. Turek}
\email[]{turek@ipm.cz}
\affiliation{Institute of Physics of Materials,
Academy of Sciences of the Czech Republic,
\v{Z}i\v{z}kova 22, CZ-616 62 Brno, Czech Republic}

\author{J. Kudrnovsk\'y}
\email[]{kudrnov@fzu.cz}
\affiliation{Institute of Physics, 
Academy of Sciences of the Czech Republic,
Na Slovance 2, CZ-182 21 Praha 8, Czech Republic}

\author{K. Carva}
\email[]{carva@karlov.mff.cuni.cz}
\affiliation{Charles University, Faculty of Mathematics and Physics,
Department of Condensed Matter Physics, 
Ke Karlovu 5, CZ-121 16 Praha 2, Czech Republic}


\date{\today}

\begin{abstract}
We present results of systematic fully relativistic first-principles
calculations of the uniaxial magnetic anisotropy energy (MAE) of a
disordered and partially ordered tetragonal Fe-Co alloy using the
coherent potential approximation (CPA).
This alloy has recently become a promising system for thin 
ferromagnetic films with a perpendicular magnetic anisotropy.
We find that existing theoretical approaches to homogeneous random
bulk Fe-Co alloys, based on a simple virtual crystal approximation
(VCA), overestimate the maximum MAE values obtained in the CPA by a
factor of four.
This pronounced difference is ascribed to the strong disorder in the
minority spin channel of real alloys, which is neglected in the VCA
and which leads to a broadening of the $d$-like eigenstates at the
Fermi energy and to the reduction of the MAE.
The ordered Fe-Co alloys with a maximum $L1_0$-like atomic long-range
order can exhibit high values of the MAE, which, however, get 
dramatically reduced by small perturbations of the perfect order.
\end{abstract}


\maketitle


\section{Introduction\label{s_intr}}

The binary ferromagnetic Fe-Co alloy has been known for a long time
as a system with the maximum spontaneous magnetization among the
transition-metal systems. \cite{r_1991_hpw}
Its basic magnetic properties, such as the concentration dependence
of the alloy magnetization (the Slater-Pauling curve) in the 
ground-state body-centered cubic (bcc) structure, have been
reproduced successfully by \emph{ab initio} electronic structure
calculations in a number of studies. 
\cite{r_1984_smb, r_1992_sej, r_1994_tkd} 
This system attracted renewed interest several years ago, after a
theoretical prediction of a giant uniaxial magnetic anisotropy
energy (MAE) of the body-centered tetragonal (bct) Fe$_{1-x}$Co$_x$
alloys. \cite{r_2004_bne}
The uniaxial MAE is defined quantitatively as the difference of total
energies for the magnetization direction parallel to the tetragonal
$a$ and $c$ axis, $K_{\rm u} = E^{(100)} - E^{(001)}$.
The calculated MAE reached high values, $K_{\rm u} \approx 
800~\mu$eV/atom, obtained however only in a narrow range of the 
Co concentration, $0.55 \le x \le 0.65$, and of the tetragonal 
strain, $1.2 \le c/a \le  1.25$. 
The combination of the high MAE with the high magnetization makes 
the bct Fe-Co system a promising material for fabrication of 
ferromagnetic thin films with a perpendicular magnetic anisotropy
(the magnetic easy axis perpendicular to the film plane), 
which might be relevant as high-density magnetic recording media.

Most of experimental realizations of this tetragonal system employed
the possibility to grow epitaxially strained Fe-Co alloy films of a
varying composition on different non-magnetic transition-metal
substrates with the face-centered cubic (fcc) structure, such as
Pd(001), Rh(001), and Ir(001). \cite{r_2008_ylt, r_2009_ypm} 
The different lattice parameters of the substrates enable one to
scan a relatively wide interval of the $c/a$-ratio of the Fe-Co
films, namely $1.13 \le c/a \le 1.24$ (note that $c/a=1$ and
$c/a=\sqrt{2}$ refer respectively to the ideal bcc and fcc lattices).
The measurements of the \emph{in situ} magneto-optical Kerr 
effect \cite{r_2009_ypm} (MOKE) confirmed the theoretical prediction
of Ref.~\onlinecite{r_2004_bne} on a qualitative level; 
in particular, a strong perpendicular magnetic anisotropy of the
films was observed for compositions and tetragonal distortions
in a rough agreement with the calculated trends of the bulk
$K_{\rm u}$.
However, systematic quantitative experimental studies of the MAE of
the films have not been performed yet; available values of the
$K_{\rm u}$ for a few selected systems 
\cite{r_2006_abw, r_2007_lfw, r_2007_wab}
are appreciably smaller than the calculated ones. 
The missing information on the MAE has partly been compensated by
measurements of X-ray magnetic circular dichroism (XMCD) spectra,
which provide orbital magnetic moments of both alloy constituents. 
\cite{r_2008_ylt}
The latter measurements, performed on the Rh(001) substrate
corresponding to $c/a \approx 1.24$, revealed enhanced orbital
moments in Fe$_{1-x}$Co$_x$ films around $x \approx 0.6$, in a close
relation to the predicted maximum of the bulk uniaxial MAE.

Nevertheless, the existing agreement between the theory and the 
experiment should be taken with caution.
The original approach of Ref.~\onlinecite{r_2004_bne} employed the
so-called virtual crystal approximation (VCA), in which the true
species of a random binary alloy are replaced by a single element
with an effective atomic number given by the alloy composition.
The VCA has been implemented in various \emph{ab initio} methods
and used for a number of alloys of neighboring elements in the
periodic table. \cite{r_1992_sej, r_2007_fth, r_2010_pch} 
Its application to the bcc Fe-Co and fcc Co-Ni systems 
\cite{r_1992_sej, r_1992_sje} yields concentration trends of the
spin magnetic moment in a very good agreement with experiment; 
\cite{r_1991_hpw} 
the orbital magnetic moments seem to be described reliably in the
VCA as well. 
However, a recent first-principles study of the random bct
Fe$_{1-x}$Co$_x$ alloys, \cite{r_2011_nsr} based on a supercell
technique applied to a few special Co concentrations 
($x=0.5$, 0.625, and 0.75), has shown that a realistic treatment
of the chemical disorder can reduce the MAE by a factor of
1.5$-$3 as compared to the simple VCA.  

The main purpose of this work is to present and discuss the results
of a systematic \emph{ab initio} calculation of the MAE for the
disordered bulk tetragonal Fe-Co systems obtained by using the
coherent potential approximation (CPA) as a basic tool of the
theory of metallic alloys. \cite{r_1982_jsf, r_2000_ag}
For homogeneous bct alloys, we compare results of the VCA and the
CPA, show the big difference between them, and identify the 
underlying physical mechanism in terms of the electronic structure.
Moreover, we investigate the effect of a complete and an incomplete
atomic long-range order on the MAE; this study is motivated by
the existing prediction of a large uniaxial MAE in stoichiometric
perfectly ordered tetragonal FeCo and FeCo$_3$ systems. 
\cite{r_2011_nsr}
Similar theoretical studies have so far been done mainly for
tetragonal FePt alloys \cite{r_2004_sor, r_2011_asc, r_2012_ks_j}
and very recently also for the FeCo alloy. \cite{r_2012_ks_e}
These works are confined to stoichiometric equiconcentration alloys;
in general, their results reveal a decrease of the MAE due to
imperfect chemical ordering.

\section{Models and computational details\label{s_mcd}}

All calculations of this study employed a bct lattice with 
the Wigner-Seitz $s$ radius equal to that of pure bcc iron, 
$s = 2.662\, a_0$ where $a_0 = 5.292\times 10^{-11}$~m denotes the
Bohr radius.
The neglect of volume relaxations on the MAE in a broad range of
lattice parameters and of alloy concentrations is justified by
a more general study of Ref.~\onlinecite{r_2011_nsr} within the
VCA. 
This structure model---the volume conserving tetragonal distortion
(the Bain path)---coincides with that used in the original study 
\cite{r_2004_bne} and in the recent study of the partially ordered
FePt alloy. \cite{r_2012_ks_j}

The effect of atomic long-range ordering has been studied for the
case of $L1_0$ (CuAu) order relevant for compositions not far from
the equiconcentration Fe$_{0.5}$Co$_{0.5}$ system.
The bct structure is partitioned in two simple tetragonal 
sublattices (alternating atomic planes perpendicular to the 
tetragonal $c$ axis).
For the Fe$_{1-x}$Co$_x$ system, an additional concentration
variable $y$ is introduced, such that $0 \le y \le \min \{x, 1-x \}$,
which defines the chemical composition of both sublattices:
Fe$_{1-x+y}$Co$_{x-y}$ in the Fe-enriched planes and
Fe$_{1-x-y}$Co$_{x+y}$ in the Co-enriched planes.
Note that the homogeneous solid solution corresponds to $y=0$,
while the other end of the $y$-interval describes the maximum
$L1_0$ order compatible with the given total Co concentration $x$.
This model is a natural generalization of the model used for
the stoichiometric FePt 
systems. \cite{r_2004_sor, r_2011_asc, r_2012_ks_j}

The eletronic structure calculations were done by means of the
tight-binding linear muffin-tin orbital (TB-LMTO) method in the
atomic sphere approximation (ASA) \cite{r_1984_aj, r_1997_tdk}
with a full inclusion of relativistic effects by solving the
one-electron Dirac equation \cite{r_1991_slg, r_1996_sdk} with
the $spd$-basis of the valence orbitals.
The effective spin-polarized potential was constructed in the
local spin-density approximation (LSDA) using the parametrization
of the exchange-correlation term according to 
Ref.~\onlinecite{r_1980_vwn}. 
The effects of alloying were treated in the CPA for all systems;
\cite{r_1997_tdk, r_1996_sdk}
for the homogeneous random alloys ($y=0$), the simple VCA was
used as well.

The reliable evaluation of the MAE in metallic systems represents
a difficult task, especially for cubic $3d$ transition metals 
owing to the weakness of the spin-orbit interaction and the high
symmetry of cubic structures. \cite{r_1995_tje, r_2000_jk}
The situation is more favorable for uniaxial systems (tetragonal,
trigonal, hexagonal) \cite{r_1992_dkb, r_2004_bej} and for systems
with $f$-electrons, see Ref.~\onlinecite{r_1998_mr} for a review.
In a number of existing studies, including the very recent ones,
\cite{r_2004_bne, r_2011_nsr, r_2012_ks_j} the magnetic force theorem
\cite{r_1987_lka} is used and the total energy difference $K_{\rm u}$
is approximated by the change in the sum of occupied valence
one-particle eigenvalues. 
In this study, we go beyond this approximation and evaluate the
MAE directly from the total energies of the fully self-consistent
solutions for both magnetization directions,
\cite{r_1995_tje, r_2011_nsr, r_2001_shh} which requires high
accuracy in total-energy calculations.
We have used uniform sampling meshes of about $10^5$ ${\bf k}$-points
for averages over the full Brillouin zone (BZ) of the bct lattices,
while about $5\times 10^4$ ${\bf k}$-points have been used for the BZ
of the simple tetragonal Bravais lattices of the $L1_0$ ordered
systems.
The total energies were converged to 0.1 $\mu$eV/atom for all
systems; this accuracy is sufficient in view of the resulting 
values of $K_{\rm u}$ (see below).

\section{Results and discussion\label{s_redi}}

\subsection{Homogeneous random alloys\label{ss_hra}}

\subsubsection{Magnetic anisotropy energy\label{sss_mae}}

\begin{figure}
\rotatebox{270}{\includegraphics[height=\columnwidth]{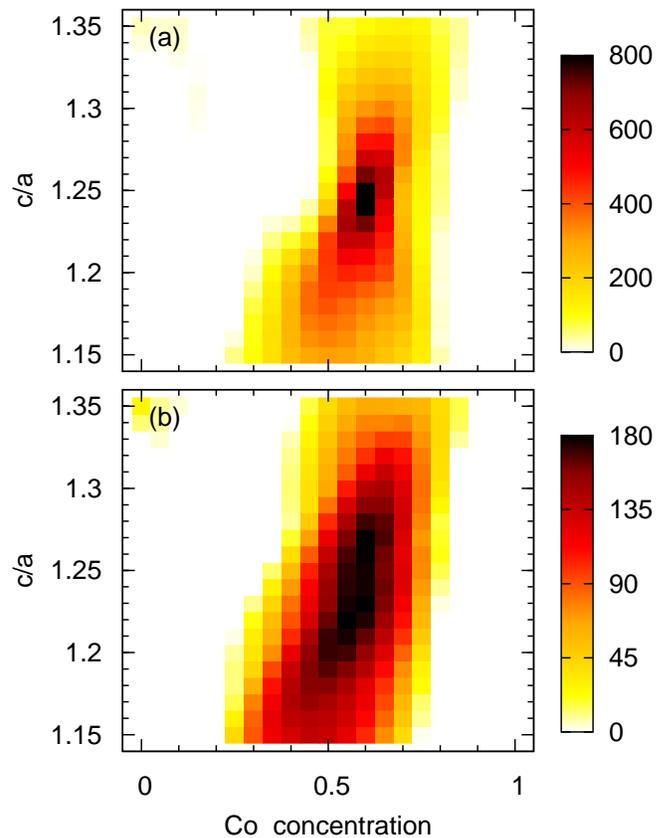}}
\caption{(Color online)
The uniaxial MAE of random bct Fe$_{1-x}$Co$_x$ alloys as
a function of the Co concentration and of the tetragonal strain
$c/a$: calculated in the VCA (a) and in the CPA (b).
Only positive values of the $K_{\rm u}$ are displayed in both plots;
the corresponding coloured scales are in $\mu$eV/atom.
\label{f_2d_mae}}
\end{figure}

The dependence of the uniaxial MAE of the random Fe$_{1-x}$Co$_x$
alloys on the chemical composition and the $c/a$-ratio, calculated
in the VCA and in the CPA, is shown in Fig.~\ref{f_2d_mae}; for
better transparency, only the positive values of the $K_{\rm u}$
are displayed in the plots, while the cases with negative
$K_{\rm u}$ have been omitted (marked by white color). 
The VCA results (Fig.~\ref{f_2d_mae}a) agree nicely with both
previous studies; \cite{r_2004_bne, r_2011_nsr} in particular,
a sharp maximum of the MAE, $K_{\rm u} \approx 809 \, \mu$eV/atom,
obtained for $x = 0.6$ and $c/a = 1.24$, is clearly visible.
The maximum MAE of $K_{\rm u} \approx 800 \, \mu$eV/atom for
$x = 0.6$ and for $c/a$ between 1.22 and 1.24 was reported in
Ref.~\onlinecite{r_2011_nsr} (with a slight sensitivity to the
particular model of the volume relaxation employed) and practically
the same values were obtained in the original VCA 
study. \cite{r_2004_bne} 
Minor quantitative differences might be ascribed to different
computational schemes employed.
The MAE in the CPA (Fig.~\ref{f_2d_mae}b) exhibits a similar trend
as in the VCA; however, the CPA maximum is significantly smaller,
$K_{\rm u} \approx 183 \, \mu$eV/atom. 
The latter value is obtained for $x = 0.6$ and $c/a = 1.23$, but
the maximum is very flat and essentially the same $K_{\rm u}$
values can be found in a broad region of $0.5 \le x \le 0.65$ and 
$1.21 \le c/a \le 1.28$, see Fig.~\ref{f_2d_mae}b.
The maximum $K_{\rm u}$ in the CPA is reduced by a factor of four
as compared to that in the VCA, which is an even stronger reduction
than that reported in Ref.~\onlinecite{r_2011_nsr}.

The small MAE values in the CPA might be used to explain the
difference between the large $K_{\rm u}$ values predicted in the VCA
and the much smaller values inferred from measurements on thin Fe-Co
films. \cite{r_2006_abw, r_2007_lfw, r_2007_wab}
Undoubtedly, the better description of the chemical disorder
in the CPA as compared to the VCA contributes to the reduction
of the MAE of prepared alloy thin films.
However, a thorough quantitative analysis of this point is impossible
at present and it could even be misleading because of the well-known
uncertainty of the LSDA to yield quantitatively correct MAEs in $3d$ 
transition-metal systems. 
For this reason, an analysis of the trends seems to be more
appropriate, see, e.g., Ref.~\onlinecite{r_2004_bej} and references
therein.

\begin{figure}
\includegraphics[width=\columnwidth]{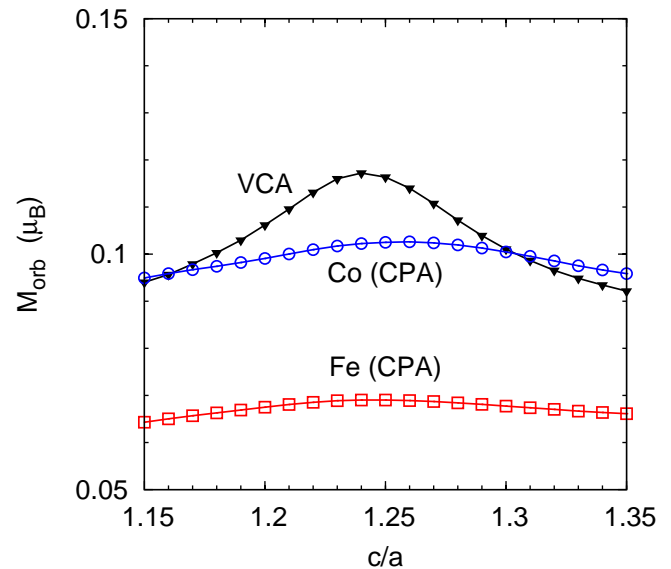}
\caption{(Color online)
The orbital magnetic moments in the random bct Fe$_{0.4}$Co$_{0.6}$
alloy with magnetization along $z$ axis as functions of the
$c/a$-ratio: the alloy orbital moment in the VCA and the 
species-resolved orbital moments in the CPA.
\label{f_orbmm}}
\end{figure}

In general, the LSDA underestimates both the MAEs and the orbital
magnetism. \cite{r_1992_sje, r_1995_tje, r_2000_he, r_2011_nsr} 
A counterexample to this rule is represented by the MAE of the
ordered FePt alloy, \cite{r_2003_sm, r_2011_asc} which
is overestimated in the LSDA probably owing to an interplay of 
intraatomic Coulomb correlations, weak exchange splitting and
strong spin-orbit interaction of Pt atoms. \cite{r_2003_sm} 
The present results for the bct Fe$_{1-x}$Co$_x$ alloys seem to
confirm the general rule as can be documented by two facts.
First, the effective MAE of strained Fe-Co thin films, given by the
difference of the bulk $K_{\rm u}$ and the magnetostatic shape
anisotropy energy $K_{\rm sh}$, comes out positive for alloys in the
vicinity of $x \approx 0.6$ and $c/a \approx 1.25$ (near the maximum
of $K_{\rm u}$ in the CPA). 
This situation would thus lead to an out-of-plane orientation of the
thin-film magnetization, in a qualitative agreement with existing
experiments. \cite{r_2009_ypm, r_2006_abw, r_2007_lfw, r_2007_wab}
Note that $K_{\rm sh} = (\mu_0/2) M^2/V$, where $\mu_0$ is the
permeability of the vacuum and where $M$ and $V$ denote, 
respectively, the alloy magnetic moment per atom and the atomic
volume.
In the present case, $M \approx 2.14 \, \mu_{\rm B}$, where
$\mu_{\rm B}$ is the Bohr magneton, one obtains 
$K_{\rm sh} \approx 132 \, \mu$eV/atom, which lies below the
bulk MAE ($K_{\rm u} \approx 180 \, \mu$eV/atom).
However, for the case of $x = 0.5$ and $c/a = 1.13$, corresponding
roughly to the Fe$_{0.5}$Co$_{0.5}$ films grown on the Pd(001)
substrate, our calculated values are $K_{\rm u} = 124 \, \mu$eV/atom,
$M \approx 2.27 \, \mu_{\rm B}$, and $K_{\rm sh} \approx 148 \, 
\mu$eV/atom, which indicates an in-plane orientation of the
magnetization, in contrast to the out-of-plane orientation observed
at low temperatures. \cite{r_2009_ypm}
Second, the calculated orbital magnetic moments in the bulk 
Fe$_{1-x}$Co$_x$ alloys, plotted in Fig.~\ref{f_orbmm} for $x=0.6$,
are appreciably smaller than the values obtained from the measured
XMCD spectra by employing the sum rules.
\cite{r_2008_ylt}
In particular, the measured values for films with $x = 0.6$ and 
$c/a = 1.24$, namely, 
$M^{\rm Fe}_{\rm orb} = 0.21 \pm 0.03\, \mu_{\rm B}$ and
$M^{\rm Co}_{\rm orb} = 0.32 \pm 0.04\, \mu_{\rm B}$, exceed their
bulk theoretical counterparts by a factor of three.
A similar relation of the experiment and the LSDA theory (quantified
roughly by a factor between 1.5 and 2) was obtained for bcc Fe-Co 
alloys. \cite{r_1992_sej, r_1992_sje, r_2000_he}
Moreover, the data in Fig.~\ref{f_orbmm} prove that the VCA and the
CPA yield qualitatively different trends of the orbital magnetism 
versus the $c/a$-ratio: the pronounced maximum at $c/a = 1.24$
in the VCA is replaced by very flat maxima of both local orbital
moments in the CPA.
This difference indicates that applicability of the VCA to all
details of magnetism of the Fe-Co system is limited, being confined
probably only to the cubic structures. \cite{r_1992_sej, r_1992_sje}

In view of the above mentioned problems to evaluate reliably the
MAEs and the orbital magnetic moments, we have not attempted to
recalculate our results by techniques going beyond the LSDA, 
\cite{r_1995_tje, r_1998_mr, r_2000_he, r_2004_bej}
but have focused on the strong difference between the VCA and the CPA
and on the role of disorder on the MAE in the tetragonal Fe-Co
systems.

\subsubsection{Electronic structure\label{sss_es}}

The giant MAE obtained in the VCA was ascribed originally to
the changes in the band structure accompanying the tetragonal
distortion of the effective FeCo crystal. \cite{r_2004_bne}
The explanation rests on a coincidence of two particular eigenvalues
at the $\Gamma$ point, which occurs in the minority spin (spin-down)
channel.
For a special alloy composition and a special $c/a$-ratio, this
coincidence takes place just at the Fermi energy corresponding
to the particular filling of the spin-down band.
The large contribution of this eigenvalue pair to the MAE can
be understood in terms of the second-order perturbation theory,
in which the effect of the weak spin-orbit interaction can be
safely included. \cite{r_1989_pb}
In this approach, the enhanced value of the $K_{\rm u}$ of the bulk
Fe-Co alloys can be explained due to the coupling between 
$d_{x^2-y^2}$ and $d_{xy}$ states mediated by the orbital momentum
operator $L_z$, see Ref.~\onlinecite{r_2004_bne} for details.
The physical origin of the perpendicular magnetic anisotropy of the
strained Fe-Co films is thus similar to that of Au covered Co 
monolayers on an Au(111) substrate. \cite{r_1996_usb}
In the latter case, however, the resulting positive MAE is caused by
the $L_z$ coupling between $d_{xz}$ and $d_{yz}$ states.

\begin{figure}
\includegraphics[width=\columnwidth]{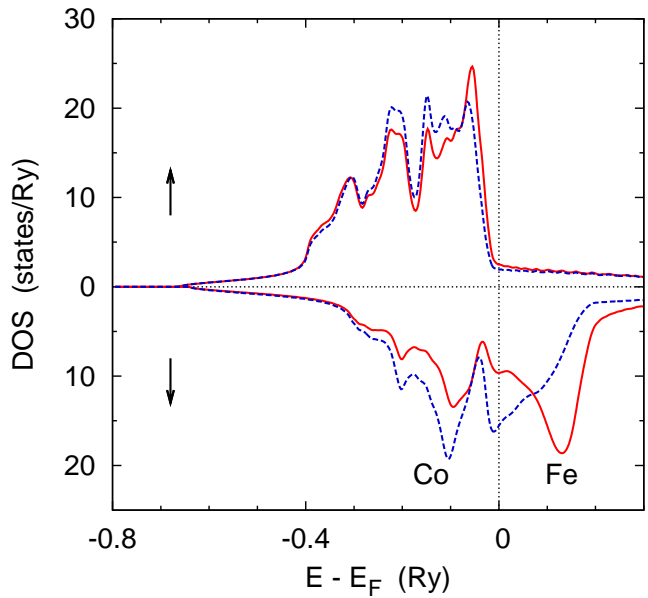}
\caption{(Color online)
The spin-polarized local densities of states of the random bct 
Fe$_{0.4}$Co$_{0.6}$ alloy with $c/a = 1.25$ as functions of the
energy.  
The dotted vertical line denotes the position of the Fermi level.
\label{f_dos}}
\end{figure}

The usual band structures are not relevant for the electronic 
structure of random alloys, for which more appropriate quantities,
such as the densities of states (DOS) and the Bloch spectral 
functions (BSF), have to be studied. 
\cite{r_1982_jsf, r_1990_pw, r_2000_ag}
Motivated by the above explanation of the giant MAE, we present
these quantities for the random bct Fe$_{1-x}$Co$_x$ alloys in the
CPA, evaluated without the spin-orbit interaction, i.e., in the 
scalar-relativistic approximation. \cite{r_1997_tdk, r_1977_kh}

The spin-polarized local DOSs are shown in Fig.~\ref{f_dos} for
the system with $x = 0.6$ and $c/a = 1.25$.
The shapes of the individual DOSs prove a very weak disorder
in the majority spin (spin-up) channel, whereas the spin-down band
exhibits a regime of a stronger scattering.
This type of spin-dependent disorder was found in the bcc Fe-Co
system a long time ago; \cite{r_1984_smb, r_1994_tkd} it is
responsible, e.g., for the observed concentration trend of the
residual resistivity. \cite{r_2009_ktt}
Note that the strong disorder is present in the spin-down $d$-band,
which is only partially occupied and which thus contributes a lot
to the alloy total energy and, consequently, to the MAE.

\begin{figure}
\includegraphics[width=\columnwidth]{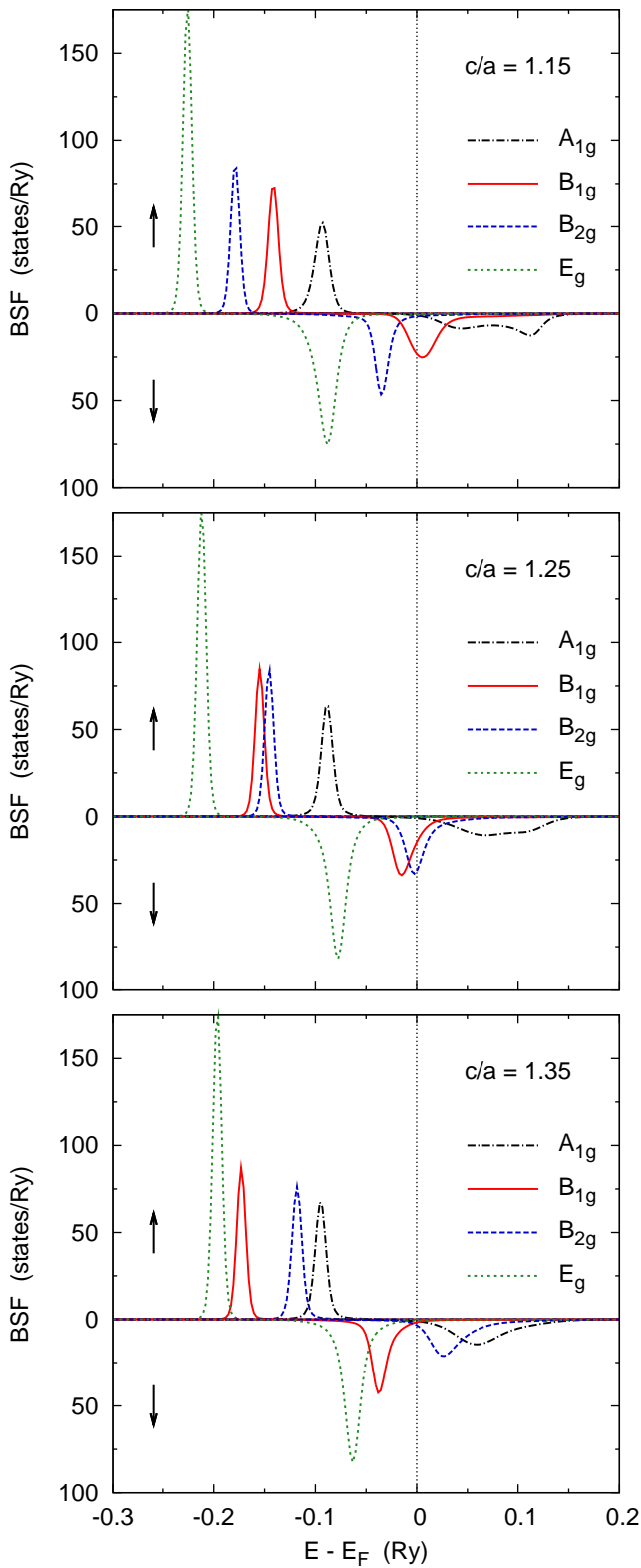}
\caption{(Color online)
The spin-polarized symmetry-resolved Bloch spectral functions of
the random bct Fe$_{0.4}$Co$_{0.6}$ alloy at ${\bf k} = \Gamma$
as functions of the energy for three values of the $c/a$-ratio.
The dotted vertical lines denote positions of the Fermi levels.
For details, see the text.
\label{f_bsf}}
\end{figure}
 
The spin-polarized BSFs at the $\Gamma$ point for the 
Fe$_{0.4}$Co$_{0.6}$ alloy with three tetragonal distortions are
plotted in Fig.~\ref{f_bsf}.
The BSFs were resolved according to the irreducible representations
of the point group $D_{\rm 4h}$ of the bct lattice. \cite{r_1960_vh}
The odd (ungerade) representations, belonging to $p$ orbitals,
have been omitted in the plots since their contribution is negligible
in the displayed energy interval. 
The relevant even (gerade) representations are: $A_{\rm 1g}$ 
(orbitals $s$ and $d_{z^2}$), $B_{\rm 1g}$ ($d_{x^2-y^2}$),
$B_{\rm 2g}$ ($d_{xy}$), and $E_{\rm g}$ ($d_{xz}$ and $d_{yz}$). 
One can see that all spin-up BSF's, consisting of narrow Lorentzian
peaks, can be interpreted as Bloch-like eigenstates with finite
lifetimes due to a weak disorder.
The spin-down BSF's above the Fermi energy (mainly $A_{\rm 1g}$-like)
reflect effects of strong disorder (non-quasiparticle behavior); the
profiles at and below the Fermi energy are Lorentzian peaks again.
However, their widths are clearly bigger than the spin-up widths,
in agreement with the spin-dependent disorder manifested in the DOS.

The maximum of the MAE in the VCA is found for an alloy with $x=0.6$
and with the tetragonal strain close to $c/a = 1.25$; its spin-down 
BSF in the CPA (Fig.~\ref{f_bsf}, middle panel) shows an analogy of
the coincidence of the $B_{\rm 1g}$ ($d_{x^2-y^2}$) and 
$B_{\rm 2g}$ ($d_{xy}$) levels at the Fermi energy.
However, the disorder-induced smearing of both peaks is at least as
big as their separation, which suppresses the contribution of this 
eigenvalue pair to the MAE in the framework of the second-order
perturbation theory. 
This feature proves that the minority-spin Fe-Co disorder is strong
enough to reduce significantly the MAE, which explains the very big
difference between the VCA and the CPA results, see 
Section~\ref{sss_mae}.

\subsection{Partially ordered alloys\label{ss_poa}}

\begin{figure}
\includegraphics[width=\columnwidth]{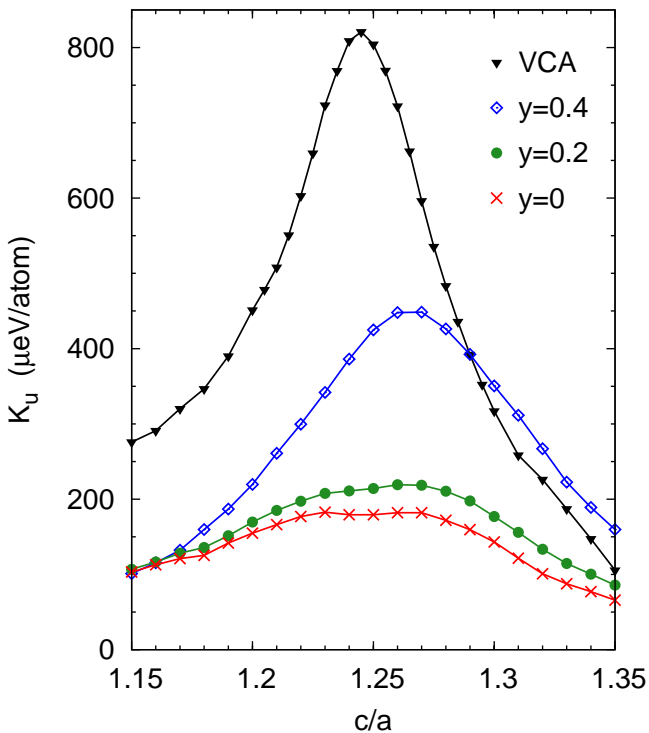}
\caption{(Color online)
The uniaxial MAE of the bct Fe$_{0.4}$Co$_{0.6}$ alloy as a function
of the tetragonal strain $c/a$ for different degrees of the 
$L1_0$-like atomic order: the maximum order ($y=0.4$, diamonds),
an intermediate order ($y=0.2$, full circles), and the complete
randomness ($y=0$, crosses), all treated in the CPA.
For a comparison, the MAE of the completely random alloy 
in the VCA is displayed as well (triangles). 
\label{f_maevsr}}
\end{figure}

The adopted model of the partially ordered tetragonal
Fe$_{1-x}$Co$_x$ alloys, 
namely, the model of two sublattices with chemical compositions
Fe$_{1-x+y}$Co$_{x-y}$ and Fe$_{1-x-y}$Co$_{x+y}$ 
(Section \ref{s_mcd}) and the use of the CPA (in contrast to 
supercell techniques)
enable one to study the MAE as a function of three continuous
variables: $x$, $y$, and $c/a$.
Such a full three-dimensional scan is, however, computationally
very demanding; we have thus confined our study to Co concentrations
close to $x=0.6$, for which the maximum MAE was found in the
homogeneous bct alloys both in the VCA and in the CPA.
The dependence of the calculated $K_{\rm u}$ on the tetragonal
distortion for $x=0.6$ is shown in Fig.~\ref{f_maevsr} for three
degrees of the $L1_0$ order.
One can see that the very flat low maximum 
$K_{\rm u} \approx 183 \, \mu$eV/atom for the completely disordered
alloy ($y=0$) treated in the CPA is replaced by a significantly 
enhanced pronounced maximum $K_{\rm u} \approx 450 \, \mu$eV/atom
obtained for the alloy with the maximum order ($y=0.4$) compatible
with the given Co concentration ($x=0.6$).
The latter maximum is about two times smaller than that in the VCA 
($K_{\rm u} \approx 809 \, \mu$eV/atom), which can be ascribed to
the fact that the system with $x=0.6$ and $y=0.4$ has one sublattice
without disorder (pure Co), whereas the other one is disordered,
containing the rest Co and all Fe atoms.
The MAE of the system with an intermediate degree of the $L1_0$ order
($y=0.2$) exhibits the maximum and the trend very close to the fully
disordered alloy.

\begin{figure}
\includegraphics[width=\columnwidth]{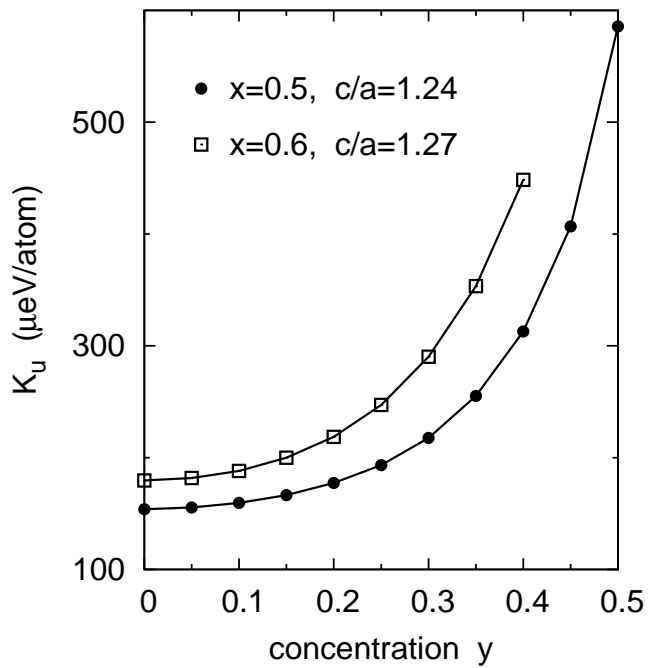}
\caption{The uniaxial MAE of two bct Fe$_{1-x}$Co$_x$ alloys
as a function of the degree of the $L1_0$-like atomic order: 
for $x=0.5$ and $c/a=1.24$ (full circles), and
for $x=0.6$ and $c/a=1.27$ (open squares).
\label{f_maevsy}}
\end{figure}

The detailed dependence of the $K_{\rm u}$ on the concentration
variable $y$ is presented in Fig.~\ref{f_maevsy} for two cases:
the stoichiometric alloy ($x=0.5$, $c/a = 1.24$) and the previous
off-stoichiometric alloy ($x=0.6$, $c/a = 1.27$).
The particular values of the $c/a$-ratio chosen in both cases
correspond to the maximum $K_{\rm u}$ value obtained for alloys
with the maximum order, i.e., for $x=0.6$, $y=0.4$ 
(Fig.~\ref{f_maevsr}) and for $x=y=0.5$.
One can see that the perfectly ordered ($y=0.5$) stoichiometric alloy
leads to a very high MAE of $K_{\rm u} \approx 580 \, \mu$eV/atom.
This value agrees well with Ref.~\onlinecite{r_2011_nsr}, where the
maximum MAE for the same $L1_0$ ordered system was obtained as 
$K_{\rm u} \approx 520 \, \mu$eV/atom.
The convex shape of both dependences in Fig.~\ref{f_maevsy} proves
that the very high MAEs can be obtained only for systems with
the maximum order; even a small amount of additional disorder is
detrimental to the MAEs and reduces them to much lower values of
the completely random alloys. \cite{r_2012_ks_e}

Strong sensitivity of the MAE to the degree of the $L1_0$ atomic
order has recently been reported for the stoichometric FePt alloy on
an fcc lattice with small tetragonal distortions. \cite{r_2012_ks_j}
The results of Ref.~\onlinecite{r_2012_ks_j} indicate that the chemical
ordering is a more important factor for high MAE values than the
tetragonal distortion.
Our results for the Fe-Co system witness that both factors are of
equal importance, see Fig.~\ref{f_maevsr}.
We believe that validity of this type of conclusions depends also
on the range of relevant variables: the tetragonal distortion
was varied over a narrower interval ($0.94 \le c/a \le 1.06$) in
Ref.~\onlinecite{r_2012_ks_j}, whereas a wider interval 
($1.15 \le c/a \le 1.35$) has been covered in our study.
Another difference between the two systems lies in the dependence
of the MAE on the $c/a$-ratio: the MAE is an ever increasing function
of $c/a$ in FePt, \cite{r_2012_ks_j} in contrast to the maxima found
for the Fe-Co systems around $c/a \approx 1.25$, see 
Fig.~\ref{f_2d_mae} and Fig.~\ref{f_maevsr}.
These trends reflect probably the different origins of the high MAE
in these systems. 
In the FePt alloys, the iron sites are responsible for strong
exchange fields and the platinum sites provide strong spin-orbit
interaction, both effects being only little dependent on the
tetragonal distortion of the lattice.
In the Fe-Co alloys, both species are featured by strong exchange
splittings and weak spin-orbit couplings; the high MAEs are heavily
based on collective properties (band structure) of the tetragonal
systems.

\section{Conclusions\label{s_conc}}

We have shown by means of first-principles LSDA calculations that
chemical disorder has a strong influence on the uniaxial MAE of the
bulk tetragonal Fe-Co systems.
First, the complete neglect of the disorder, inherent to the simple
VCA, overestimates the MAE by a large factor, whereas the more
sophisticated CPA leads to a drastic reduction of the MAE.
The latter low MAEs are quite close to the magnetic shape anisotropy
energy and the resulting estimated stability of the perpendicular
magnetic anisotropy of thin strained Fe-Co films comes out probably
smaller than the measured one.
Second, the disorder-induced reduction of the MAE is due to the
strong scattering regime in the minority-spin channel, which
smears the Bloch-like eigenstates at the Fermi energy.
Third, an analysis of the calculated orbital magnetic moments points
to a non-negligible underestimation of the orbital magnetism and,
most probably, of the MAEs by the LSDA.
Fourth, the $L1_0$-like atomic long-range order leads to an
enhancement of the MAE values. 
The most pronounced enhancement is obtained for the maximum degree
of the order, while imperfect ordering reduces the MAE very 
rapidly as compared to the perfectly ordered systems. 
Similar correlations between the atomic order and the MAE have
recently been found in the FePt alloy 
\cite{r_2004_sor, r_2011_asc, r_2012_ks_j} 
and observed in artificially synthesized FeNi 
films. \cite{r_2012_kmk}
We believe that this interplay should be taken into account in a
future search of advanced materials for high-density magnetic
recording. 

\begin{acknowledgments}
The work was supported by the Czech Science Foundation 
(Grant No.\ P204/11/1228).
\end{acknowledgments}


\providecommand{\noopsort}[1]{}\providecommand{\singleletter}[1]{#1}%

\end{document}